\documentclass{pasj00}
\draft
\begin{document}
\SetRunningHead{Kim et al.}{Systematic Microwave Source Motions}

\title{Systematic Microwave Source Motions along Flare-arcade Observed by Nobeyama Radioheliograph and AIA/SDO}

\author{Sujin \textsc{Kim}}
\affil{Nobeyama Solar Radio Observatory / NAOJ, Nagano, 384-1305,
Japan} \email{sujin.kim@nao.ac.jp}
\author{Satoshi \textsc{Masuda}}
\affil{STElab / Nagoya University, Nagoya, 464-8601
Japan}\email{masuda@stelab.nagoya-u.ac.jp}
\author{Kiyoto {\sc Shibasaki}}
\affil{Nobeyama Solar Radio Observatory / NAOJ, Nagano, 384-1305,
Japan}\email{shibasaki.kiyoto@nao.ac.jp} \and
\author{Su-Chan {\sc Bong}}
\affil{Korea Astronomy and Space Science Institute, Daejeon,
305-348, Republic of Korea}\email{scbong@kasi.re.kr}


\maketitle

\begin{abstract}
We found systematic microwave source motions along a flare-arcade
using Nobeyama Radioheliograph (NoRH) 17 GHz images. The motions
were associated with a X-class disk flare which occurred on 15th
February 2011. For this study, we also used EUV images from
Atmospheric Imaging Assembly (AIA) and magnetograms from
Helioseismic and Magnetic Imager (HMI) onboard Solar Dynamics
Observatory, and multi-channel microwave data from Nobeyama
Radiopolarimeters (NoRP) and Korean Solar Radio Burst Locator
(KSRBL). We traced centroids of the microwave source observed by
NoRH 17 GHz during the flare and found two episodes of the motion
with several facts: 1) The microwave source moved systematically
along the flare-arcade, which was observed by the AIA 94 \AA\, in
a direction parallel to the neutral line. 2) The period of each
episode was 5 min and 14 min, respectively. 3) Estimated parallel
speed was 34 km s$^{-1}$ for the first episode and 22 km s$^{-1}$
for the second episode. The spectral slope of microwave flux above
10 GHz obtained by NoRP and KSRBL was negative for both episodes,
and for the last phase of the second episodes, it was flat with
the flux of 150 sfu. The negative spectrum and the flat with high
flux indicate that the gyrosynchrotron emission from accelerated
electrons was dominant during the source motions. The sequential
images from the AIA 304 \AA\ and 94 \AA\ channels revealed that
there were successive plasma eruptions and each eruption was
initiated just before the start time of the microwave sources
motion. Based on the results, we suggest that the microwave source
motion manifests the displacement of the particle acceleration
site caused by plasma eruptions.
\end{abstract}

\section{Introduction}
Emissions generated by high energy particles during solar flares
provide key information of accelerations, injections, and
trap-precipitations in the flaring structures. In practice, during
flares, the microwave in high frequency regime can capture
optically thin gyrosynchrotron emission from accelerated
electrons. Nobeyama Radioheliograph \citep{nak94} has been taking
images of the full sun, including flares as well as the quiet sun,
at 17 GHz since 1992 and at 34 GHz since 1994. Using these
microwave data, many authors have studied the distribution of
accelerated electrons and its dynamics in flaring loops
\citep{kun01, mel02, yok02, hua09, rez10, asa13}. Simultaneously,
particle kinematics in flare loops have been established by models
and simulations based on NoRH observations \citep{mel02, mel05,
min08, min10}.

When flares occur, accelerated high energy electrons are generally
revealed in microwaves by gyrosynchrotron emission in the magnetic
loops and in hard X-ray (HXR) by bremsstrahlung at the footpoints
of the loops \citep{shi95}. Thus, a position of a source observed
in the HXR and microwaves can manifest a macroscopic view of flare
dynamics as a direct indicator of the change of the magnetic field
configuration of flaring region. And a motion of the source
position can be regarded as a displacement of a site where the
injection and the acceleration take place. The HXR bremsstrahlung
from energetic electrons at the footpoints of the magnetic loop
has shown various patterns of source motions \citep{kru03, bog05,
gri05, liu09, liu11}. \citet{bog05} studied the motions of two
footpoints linked to opposite sides of the magnetic neutral line
(NL) during 31 flares and categorized the major patterns of the
motions into three types: perpendicular, anti-parallel, and
parallel to the NL. They interpreted the former two types as a
standard reconnection model with simple or sheared overlying
magnetic field and the last type was speculated as a chromospheric
signature of a displacement of the particle acceleration region in
the corona. \textbf{Similarly to HXR footpoints motion,
\citet{rez10} found that one footpoint of microwave flare loop
moved while the other was stationary. The authors interpreted it
as an unshearing motion of the magnetic loops resulting from the
energy releasing process during the flare. Recently, Fleishman et
al. (2013) reported the displacement of the centroid of microwave
source during the flare. It appeared in the edge of the HXR
source, which was located near the top of the thermal loop, during
the impulsive phase, and then moved to the center of the HXR
source during the decay phase. They speculated that the microwave
emission dominated the acceleration site of electrons during the
early phase of the flare and then dominated the trapping site of
accelerated electrons. This centroid position and motion of
microwave source can give a clue to solve dynamics of high energy
particles during the flare through showing direct position where
the particles are accelerated and trapped.}

In this paper, for the first time, we present systematic motions
of the microwave source centroid along the flare-arcade system.
The motion recurred once again in the similar way with the first
episode during the flare. It was investigated using available high
time cadence multi-wavelength data: microwave imaging data from
NoRH 17 GHz, and EUV imaging data from 304 and 94 \AA\ channels of
Atmospheric Imaging Assembly (AIA: \cite{lem12}) and magnetograms
of Helioseismic and Magnetic Imager (HMI: \cite{sch12}) onboard
Solar Dynamics Observatory. Also, we have examined HXR sources
observed by RHESSI, but they showed no systematic motion. Thus, we
focused on the microwave source motion in the observations. The
discrepancy between the motions in HXRs and microwaves is
discussed in \S3. In \S2, we describe the morphological evolution
in EUV, the source motions in microwave, and the microwave spectra
during two episodes of the motions. In \S3, we discuss possible
cause of the microwave source motions during the flare.

\section{Observations}
We have examined a X2.2 flare which occurred on 15 February 2011.
The flare took place in Active Region 11158 which was located near
the disk center, S20W11. Figure 1 shows the light curves of fluxes
of GOES X-rays (top), NoRP 17 and 35 GHz (middle), and RHESSI
25-100 keV (bottom). The microwave flare started at 01:48:19 UT
and peaked at 01:55:20 UT. There was another peak before main peak
at 01:53:30 UT. These two peaks are separated by the period of a
sudden flux drop. The flux curve of HXR in the bottom panel of
Figure 1 also exhibits consistent time profile with the microwave
flux, which has a drastic drop between two peaks. It suggests that
there were two distinct energy release processes resulting in two
peaks in the time profile of the flux. In this section, we present
plasma eruptions and microwave source motions in which overall
observations are intimately related with each other in timing
aspect as appears in the flux curves. And we have examined the
microwave spectra and flux value to know the emission mechanism of
moving sources.

\subsection{Plasma Eruptions}
EUV multi-channels of AIA \textbf{aim for} investigation of
multi-thermal plasma in the solar atmosphere from the chromosphere
to the corona. In order to look at the emission from the
chromosphere and the corona simultaneously, we have examined two
EUV channels. One is 304 \AA\ channel, which contains a
contribution from He\,{\sc II} formed at 0.05 MK, and the other is
94 \AA\ channel, which contains lines from Fe\,{\sc XVIII}, formed
at 7 MK \citep{lem12}. The AIA data were taken with a spatial
resolution of 1.2 \arcsec\ every 12 seconds. Figure 2 shows
sequential images of two channels from top to bottom during the
flare. The 304 \AA\ channel exhibits the formation and the
development of the two-ribbon at the chromospheric footpoints. The
neutral line (a black line in Figure 2d) is located on the center
of the two-ribbon. The neutral line, where the positive polarity
and the negative polarity are encountered, was determined by
magnetogram obtained from HMI (see a yellow line in Figure 4). At
the same time, the 94 \AA\ channel shows the formation of
flare-arcade loops filled with hot plasma during the flaring
process. The sequential images of both channels reveal that there
were two successive eruptions during impulsive phase of the flare.
First eruption (1st ER) is shown indirectly in 94 \AA\ images as a
rapid radial expansion of ambient coronal loops. Figure 2e and 2f
show one of them which expands along white arrows. The ambient
coronal loops began to expand just before the flare start time,
and finally, they shrunk and/or disappeared. Second eruption (2nd
ER) appeared in the 304 \AA\ channel as a bulk plasma eruption. It
was initiated in the north-east site of the two-ribbon system
around 01:53 UT and erupted along the direction of a white arrow
in Figure 2c. Coronal mass ejections (CME) observed by Large Angle
Spectroscopic Coronagraph (LASCO: \cite{bru95}) onboard Solar and
Heliospheric Observatory (SOHO: \cite{dom95}) have been reported
in on-line CME
catalog\footnote{http://cdaw.gsfc.nasa.gov/CME\_list}. According
to the catalog, the eruption related to our flare was developed
into a halo CME. Considering the radial expansion and timing of
the first eruption, it is plausible that it results in the
observed halo CME. On the other hands, it seems that the second
eruption could not be seen in the LASCO images by screening of
huge and strong emission of the first CME and/or by a failure of
escape from the solar atmosphere.

\subsection{Microwave Source Motions}

We have investigated the source motions using NoRH 17 GHz images
with 1 second time cadence. Synthesized images of 17 GHz have a
spatial resolution of 10 \arcsec. We determined the source
centroid as a center of 95 \% level of maximum brightness
temperature ($T_B$). As the result, we found smooth trajectory of
the source motions. Figure 3 shows two episodes of the microwave
source motion with contours of $T_{B}$ at 17 GHz on the AIA 94
\AA\ image. For the first episode (here after EP1) in the left
panel of Figure 3, a compact microwave source appears on the east
site of the flaring region at 01:48:19 UT and then migrates toward
the north-west direction along the flare arcade until 01:53:34 UT.
The period of the motion is around 5 minutes. The second episode
(here after EP2) in the right panel of figure 3 begins near the
start position of EP1 at 01:55:30 UT and the source moves toward
the north-west direction along the flare arcade until 02:10:00 UT.
The period of the motion is around 14 minutes. The source
centroids during each episode are shown in the bottom panels of
Figure 3, with colors coded by time. The uniformly distributed
color manifests that the sequential motion progressed
systematically. Note that the source positions are located near or
on the top of the flare-arcade seen in the 94 \AA\ image. If the
maximum of microwave emission are due to gyrosynchrotron emission
of accelerated electrons during two episodes, then the results
imply that the accelerated particles are generated at the loop-top
or high in the corona and injected into the loop-top. Thus, one
can speculate that the displacement of the acceleration site comes
into view as a microwave source motion during the flare.
In order to investigate the source motion in detail, we plot all
of the source centroid on the magnetogram obtained by HMI (Figure
4). Source positions are elongated parallel to the neutral line
which is denoted by a yellow line in Figure 4. Figure 5 shows the
variation of the source position with time in the direction
parallel to the NL (top) and perpendicular to the NL (bottom). The
parallel motion shows consistent propagation along the NL and its
speeds estimated by the linear fit was 37 km s$^{-1}$ for EP1 and
22 km s$^{-1}$ for EP2. Meantime, the perpendicular motion is
converging toward the NL with a speed of -9 km s$^{-1}$ for EP1
and -3 km s$^{-1}$ for EP2. Considering the center of the
microwave sources are located near the loop-top region (bottom
panels of Figure 3), the converging motion toward the NL may
reflect that the loop-top position of flare loops moves toward the
NL.

\subsection{Microwave Spectra}

Nobeyama Radiopolarimeters (NoRP: \cite{nak85}) has multi-channels
of 1, 2, 3.75, 9.4, 17, 35, and 80 GHz. It detects total flux from
the sun with a time cadence of 0.1 s. During flares, microwave
spectra obtained by NoRP can be used to identify the emission
mechanism in high frequency regime, gyroshynchrotron emission from
mildly relativistic electrons or free-free emission due to thermal
electrons (see \cite{dul85}). In this study, low frequency
channels of 1 and 2 GHz were not used because they didn't follow
gyrosynchrotron nor thermal free-free emission. 80 GHz also wasn't
used due to no detection. In addition to the NoRP data, we have
examined continuous spectra using Korean Solar Radio Burst Locator
(KSRBL: \cite{dou09}) which is a radio spectrometer covering
0.245, 0.410, and 0.5---18 GHz band with 1 second time cadence and
1 MHz frequency resolution. \textbf{Due to the characteristics of
the spiral feed in use, it observes only the right-hand circular
polarization (RCP) and the observed spectra are modulated in a
regular pattern depending on the source location. As microwave
spectra are expected to follow a power law \citep{dul85}, the
modulation pattern can be removed by minimizing the average
deviation from a smooth fit of the demodulated spectra, and it
results in the ability to locate the source position within 2
\arcmin. To compare with the NoRP data, we matched the flux of the
KSRBL at quiet time to the Stokes I flux of US Air Force Radio
Solar Telescope Network (RSTN: \cite{gui81}) assuming no
polarization.} Since strong interferences appeared below 9.5 GHz,
we only used the frequency band above this frequency.

Figure 6 shows microwave spectra obtained by NoRP four channels,
3.75, 9.4, 17, and 35 GHz. Each spectrum, from the top to bottom,
was obtained near the start time and the end time of each episode.
During EP1 (Figure 6a and 6b), microwave spectra demonstrate that
the gyrosynchrotron emission is consistently dominant: decreasing
flux with frequency in high frequency regime with a turnover
frequency around 9.4 GHz (vertical dashed line in Figure 6). The
KSRBL continuous spectra is in alignment with the straight line
connecting fluxes at 9.4 GHz to 17 GHz of the NoRP. It confirms
that the turnover frequency is not higher than 10 GHz. On the
other hands, during EP2, the gyrosynchrotron emission is dominant
only during initial phase (Figure 6c) and then the spectra become
flat with constant flux of $\sim 150$ sfu in high frequency regime
(Figure 6d). One should note that 150 sfu is too high value just
to be generated by thermal free-free process, but rather there
would be significant contribution from gyrosynchrotron emission.
\citet{whi11} have presented that GOES X1 flare produces an order
of 10--50 sfu of thermal emission from optically thin flare plasma
above 10 GHz. We estimated expected radio thermal emission based
on the radiative transfer equation for thermal free-free emission
of optically thin plasma \citep{dul85} using physical quantities
of plasma temperature, 15 MK, and emission measure, $1 \times
10^{49}$ cm$^{-3}$, which were derived from GOES two-channels (see
\cite{whi05}). As the result, we found that the contribution of
microwave thermal emission was around 90 sfu which is much less
than the observed flux. Considering the compact source for EP1 and
broad source covering flare arcade for EP2, thermal emission were
enhanced during initial phase of nonthermal flaring process so
that it might come into view as the broad source covering large
area of flaring region during EP2.

\section{Summary and Discussion}

We have investigated microwave source motions during X-class flare
which occurred on 15 February 2011 using NoRH 17 GHz, AIA two
channels of 304 and 94 \AA, and HMI magnetorgram. As the results,
we found two episodes of microwave source motion along the
flare-arcade and the neutral line. All of the source positions
were located near the top of the flare loops constituting the
flare-arcade. Interestingly, two episodes of source motion started
just after the initiation of two successive plasma eruptions
revealed by sequential images of the 304 and the 94 \AA. It
suggests that the plasma eruptions may trigger the microwave
source motion and may govern the systematic motion along the
flare-arcade and the NL.

The standard model of the flare (e.g. \cite{hir74,shi95,for96})
describes the energy release process with several key sequential
activities: 1) Rising flux rope system, 2) bi-directional magnetic
field lines closing by inward pressure in between them where the
flux rope sweeps through, 3) reconnection and then release of
accelerated particles toward the rising flux rope and the flare
loops. When the accelerated particles are injected and trapped
into newly formed magnetic loops by reconnection, gyrosynchrotron
emission from them is readily generated in microwave and its
position would be on or near the loop-top region. We found that
the eruptions, which are a manifestation of rising flux ropes,
appeared in the 304 and 94 \AA\ images (Figure 2) and halo CME was
followed in SOHO/LASCO. Hence, in the picture of the standard
model, the source position on the loop-top implies that the
accelerated electrons were injected into the loop-top directly
from high corona \citep{mel02} or the fast contraction of newly
formed loop by reconnection accelerated electrons in the loop-top
up to relativistic energies \citep{som97,bog05}. In fact, the
loop-top concentration of microwave emission in high frequency has
been frequently observed by NoRH \citep{kun01, mel02, mel05,
min10, asa13}.

The most plausible scenario to interpret the systematic microwave
source motions is ``asymmetric filament eruptions'' \citep{tri06,
liu09, liu11}. In practice, the hard X-ray footpoints motion
parallel to NL in the same direction for both footpoints could
take place when the filament erupts asymmetrically: One end of the
filament rises up by unstable physical condition while the other
end anchors to the photosphere and it can lead to a sequential
magnetic reconnection in overlying loops aligned with the neutral
line (see the illustration of whipping like eruption in Liu et al.
2009). \citet{liu11} demonstrated above scenario with the
observations of the asymmetric CME eruption and the HXR source
motions parallel to the neutral line. Indeed, the microwave source
position located in the corona is closer to the acceleration site
than HXR source position located on the chromosphere. Thus, the
microwave source motion can be a direct indicator of direction of
moving acceleration site. On the other hands, recently,
\citet{nak11} proposed a model that a slow magnetioacoustic wave
triggered by an elementary burst may cause another elementary
burst in coronal magnetic arcade and it propagates along this
arcade. In this model, the elementary burst above or in the
loop-top of one edge of the arcade system is enough to produce the
motion of high energy emission at the footpoint and near the
loop-top.

Note that the high energy emission appears simultaneously in HXR
and microwave by injection and precipitation of accelerated
particles, and one can expect that their motion should be
consistent in the direction and/or its systematic progress.
However, in this event, we found that HXR source motion didn't
follow microwave source motion. Furthermore its position was not
close to the footpoints where the chromospheric brightening
appeared in AIA 304 \AA. Such inconsistency can be explained by a
presumption as follows: Too strong B field of loops prevents the
effective precipitation of trapped electrons by magnetic mirror
effect when they are close to footpoints of field \citep{whi11}.
At the same time with the above effect, the complex magnetic field
configuration of nonthermal loops, unlike well aligned thermal
flare-arcade, may lead to unsystematic motion in hard X-rays.
Nonetheless, all the summed results strongly support that our
finding of microwave source motion manifests the displacement of
the site where accelerated electrons are injected and trapped.

This is the first report of the systematic motion of the microwave
source centroid. Even so, it seems to be a common phenomenon which
takes place together with the filament/plasma eruption. In fact,
hot thermal plasma generated by energy-release process has long
cooling time-scale, around several minutes at least, while
nonthermal plasma has short lifetime, within 1 seconds for HXR
\citep{kip95} and \textbf{1$\sim$100 seconds for microwaves
\citep{min08}}. Thus, nonthermal emission of microwave and hard
X-rays is essential to detect and trace the energy-release site
than thermal emission of soft X-rays and EUV. The available time
cadence of NoRH microwave image is 1 second for steady-mode and
0.1 second for event-mode, while RHESSI HXR needs 4 seconds
\citep{gri05} and/or 8 seconds \citep{kru03} average to construct
image successfully. In turn, the microwave imaging data from NoRH
can effectively trace the time-series of the motion of flare
energy-release site even for very short period motion less than 1
second. In this study, the high time cadence data set from NoRH
and AIA allowed to examine the source motion and the evolution of
the thermal structure in detail during the flare. Thus, the
statistical study of them using these data set may provide
information on physical conditions crucial for the initiation of
the filament/plasma eruption at the position where the microwave
source motion starts.

\ The authors thanks to the referee for helpful comments. This
work was carried out by the joint research program of the
Solar-Terrestrial Environment Laboratory, Nagoya University. The
HMI and AIA data have been used courtesy of NASA/SDO and the AIA,
EVE, and HMI science teams. RHESSI is a NASA Small Explorer
Mission.


\begin{figure*}
  \begin{center}
    \FigureFile(90mm,90mm){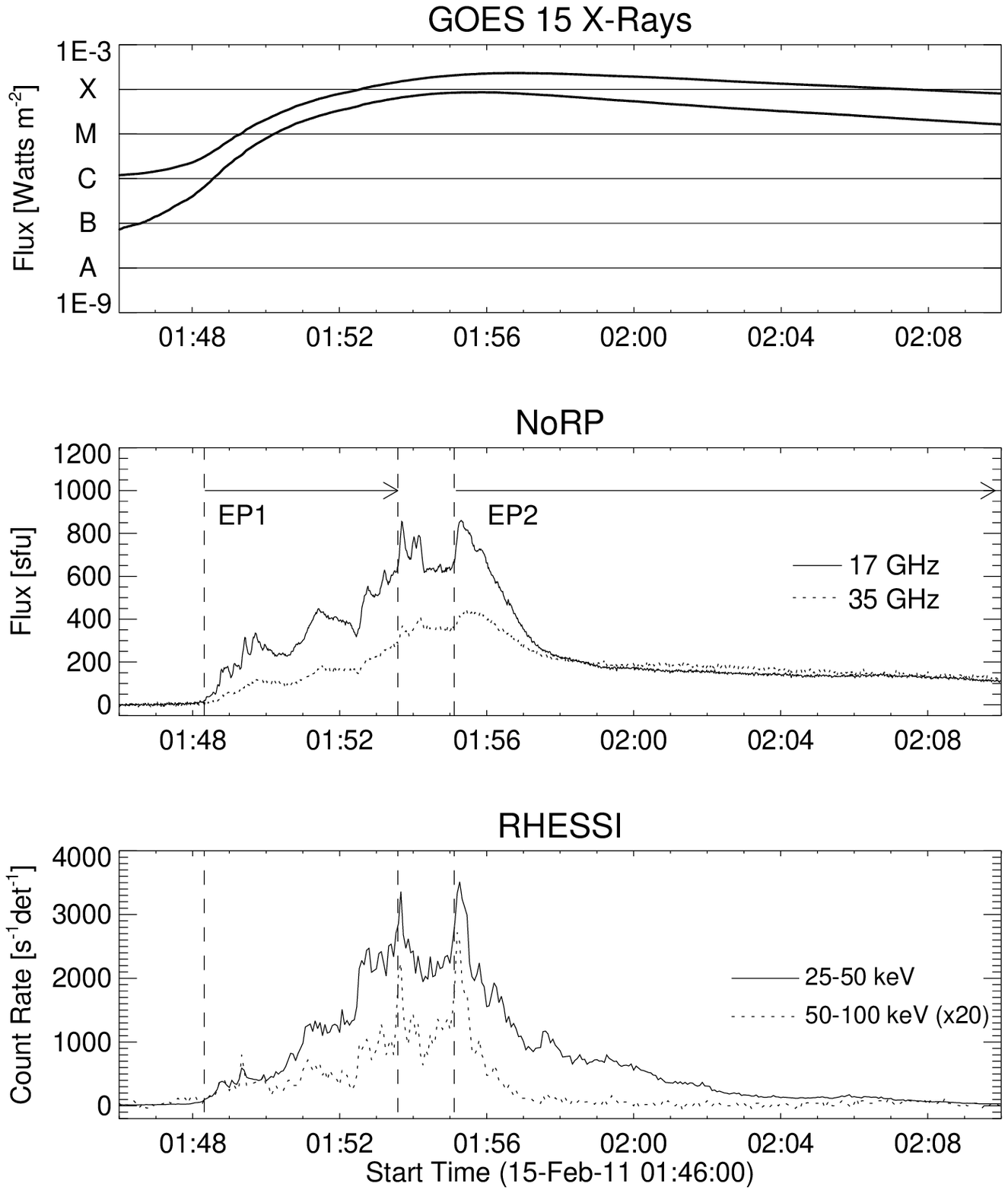}
  \end{center}
  \caption{Time profiles of fluxes from GOES X-rays (top), NoRP 17 and 35 GHz (middle),
   and RHESSI 25-50 and 50-100 keV (bottom). Vertical dashed lines
   are start and end times of two episodes (EP1 and EP2) of microwave source
motion}\label{Figure1}
\end{figure*}

\begin{figure}
  \begin{center}
    \FigureFile(90mm,90mm){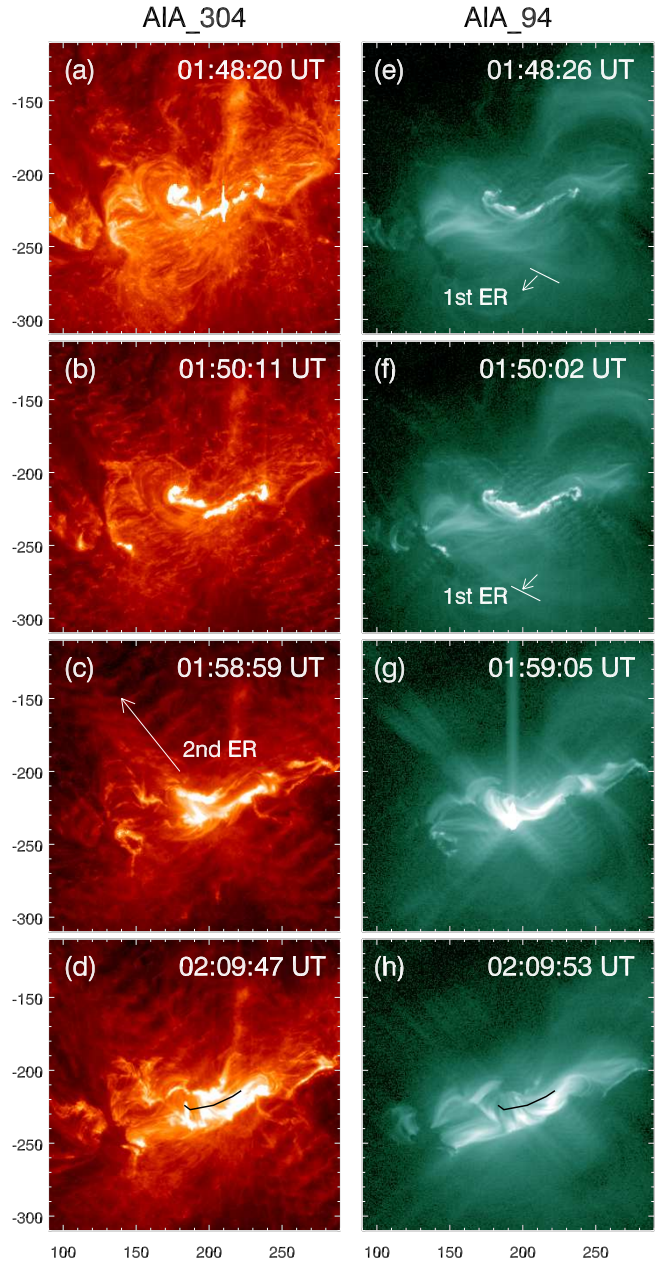}
  \end{center}
  \caption{Sequential images from the AIA two channels, 304 \AA\ (left column) and 94 \AA\ (right
  column).The 94 \AA\ channel exhibits the manifestation of the first
  eruption (1st ER) in (e) and (f): The edge of a bundle of
  coronal loops, which is marked by solid lines, expands along a white arrow. The
  304 \AA\ channel exhibits the second eruption (2nd ER) in (c): bulk of plasma erupts along
  a white arrow. The neutral line determined by the HMI magnetogram (Figure 4) is superimposed
  on the bottom pannels as a black line}\label{Figure2}
\end{figure}

\begin{figure*}
  \begin{center}
    \FigureFile(150mm,150mm){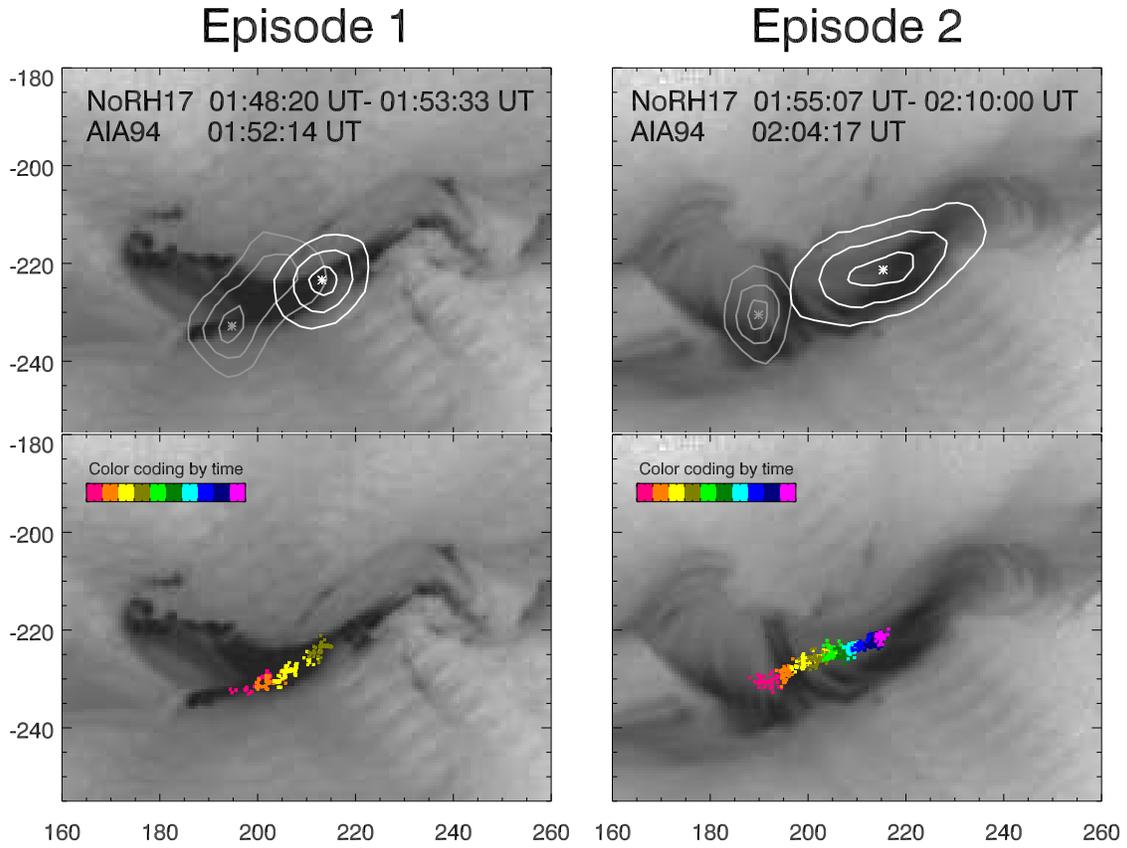}
  \end{center}
  \caption{\emph{Top panels:} NoRH 17 GHz contours on AIA 94 \AA\ image
  for two episodes of the microwave source motion, episode1 (left) and episode2 (right).
  The levels of contour are 70, 90, and 95 \% of the maximum $T_B$. Gray and white contours
  were obtained at the start and the end time of the source motion, respectively.
  \emph{Bottom panels:} Overall positions of microwave source
  centroid during episode1 (left) and episode2 (right). Colors of each symbol are coded by time.
  Each color contains 90 seconds and the colorbar covers 15 minutes with 10 colors.}\label{Figure3}
\end{figure*}

\begin{figure}
  \begin{center}
    \FigureFile(70mm,70mm){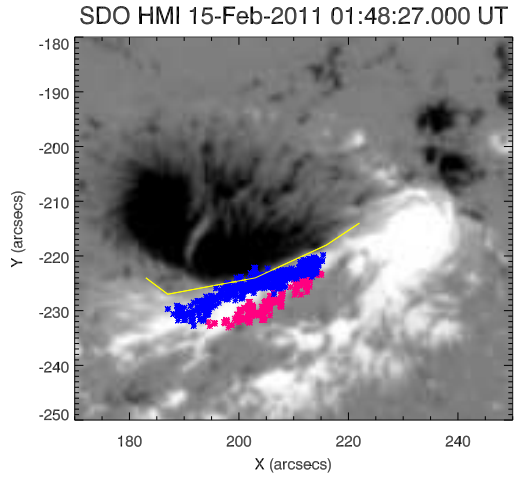}
  \end{center}
  \caption{Microwave source centroids for EP1 (red) and EP2 (blue) superposed on
  the HMI magnetogram. The neutral line is denoted by a yellow line.}\label{Figure4}
\end{figure}

\begin{figure}
  \begin{center}
    \FigureFile(70mm,70mm){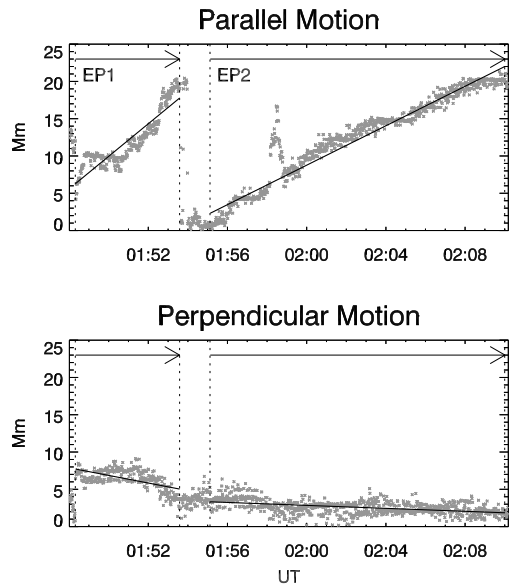}
  \end{center}
  \caption{Time evolution of the source positions parallel to the NL (top)
  and perpendicular to the NL (bottom). The velocity of motions was derived by
  linear fit of the scattered plots of positions (solid lines).}\label{Figure5}
\end{figure}

\begin{figure}
  \begin{center}
    \FigureFile(80mm,80mm){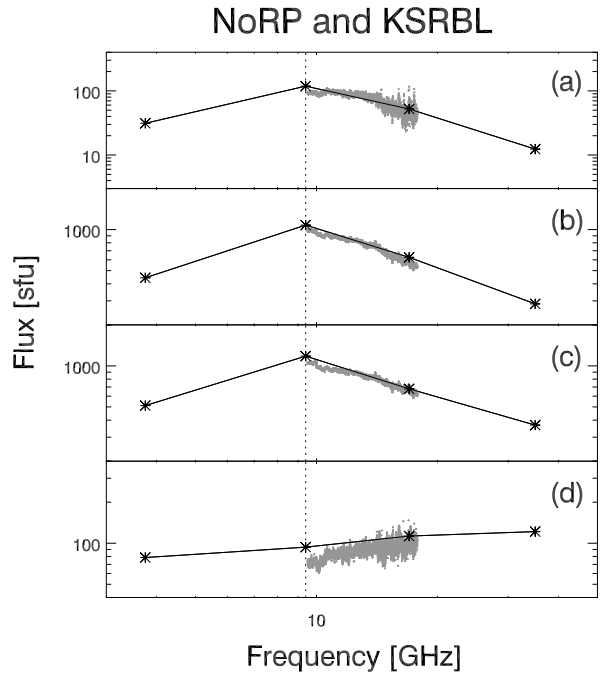}
  \end{center}
  \caption{Microwave spectra obtained by NoRP four channels of 3.75, 9.4, 17,
  and 35 GHz (asterisks) and by KSRBL continuous spectra in the range of 9.5--18
  GHz with 1 MHz resolution (scatter plot with gray color) during the flare.
  (a) and (b) were obtained at 01:48:34 UT and at 01:53:34 UT during episode 1. (c) and (d)
  were obtained at 01:55:09 UT and at 02:09:59 during episode 2.
  Vertical dashed line marks 9.4 GHz of NoRP which roughly indicates
  the turnover frequency.}\label{Figure6}
\end{figure}

\end{document}